# An Optimal Framework for Residential Load Aggregators

Qinran Hu, *Student Member, IEEE*, Fangxing Li, *Senior Member, IEEE*

*Abstract*—Due to the development of intelligent demand-side management with automatic control, distributed populations of large residential loads, such as air conditioners (ACs) and electrical water heaters (EWHs), have the opportunities to provide effective demand-side ancillary services for load serving entities (LSEs) to reduce the emissions and network operating costs. Most present approaches are restricted to 1) the scenarios involving with efficiently scheduling the large number of appliances in real time; 2) the issues about evaluating the individual residents' contributions towards participating demand response (DR) program, and fairly distributing the rewards; and 3) the concerns on preforming cost-effective LSEs' demand reduction request (DRR) with minimal rewards costs while not effecting residents' living comfortableness. Therefore, this paper presents an optimal framework for residential load aggregators (RLAs) which helps solve the problems mentioned above. Under this framework, RLAs are able to realize the DRR for LSEs to generate optimal control strategies over residential appliances quickly and efficiently. To residents, the framework is designed with probabilistic model of comfortableness, which minimizes the impact of DR program to their daily life. To LSEs, the framework helps minimize the total reward costs of performing DRRs. Moreover, the framework fairly and strategically distributes the financial rewards to residents, which may stimulate the potential capability of loads optimized and controlled by RLAs in demand side management. The proposed framework has been validated on several numerical case studies.

NOMENCLATURE

| | |
|---|---|
| $n$ | Number of households under one RLA. |
| $R_1$ | Level 1 reward rate, cents/(kW·5min). |
| $R_2$ | Level 2 reward rate, cents/(kW·5min). |
| $R_3$ | Level 3 reward rate, cents/(kW·5min). |
| $T_{RM,i}$ | Room temperature for resident $i$, ℉. |
| $T0_{RM,i}$ | Initial room temperature for resident $i$, ℉. |
| $T_{L,i}$ | Low room temperature threshold for resident $i$, ℉. |
| $T_{H,i}$ | High room temperature threshold for resident $i$, ℉. |
| $PA_i$ | AC power rate of resident $i$, kW. |
| $SA_i$ | Operating status of the AC of resident $i$, ON/OFF. |
| $AE_i$ | Effect of the AC of resident $i$, ℉/kW. |
| $LR_{RM,i}$ | Room temperature loss rate of resident $i$. |
| $RWR_{A,i}$ | AC reward rate for resident $i$, cents/(kW·5min). |
| $T_{T,i}$ | EWH tank temperature of resident $i$, ℉. |
| $T0_{T,i}$ | EWH initial tank temperature of resident $i$, ℉. |
| $T_{TL,i}$ | Low tank temperature threshold of resident $i$, ℉. |
| $T_{TH,i}$ | High tank temperature threshold of resident $i$, ℉. |
| $PE_i$ | EWH power rate of resident $i$, kW. |
| $SA_i$ | EWH operating status of resident $i$, ON/OFF. |
| $E_i$ | Effect of the EWH for resident $i$, ℉/kW. |
| $LR_{T,i}$ | Tank temperature loss rate for resident $i$. |
| $RWR_{E,i}$ | EWH reward rate for resident $i$, cents/kW·5min. |
| $Cop_i$ | Whether resident $i$ is willing to compromise the appliances operating beyond their comfortable interval, ON/OFF. |
| $CM_i$ | Comfortable margin for resident $i$. |
| $TA_i$ | Ambient temperature for resident $i$. |
| $TDR$ | Total demand secluded to reduce, kW. |
| $D$ | Amount of demand reduction required, kW. |
| $\delta$ | Parameter associated with demand reduction accuracy relaxation. |
| $RW_i$ | Total financial rewards for resident $i$, $. |
| $w$ | Weight of comfortable margin. |
| $\mu_i, \upsilon_i$ | Auxiliary binary variables for converting the optimization problem. |

I. INTRODUCTION

The development of communication and sensing technologies provide more advanced platforms for electricity consumers and suppliers to interact with each other. According to the monthly energy review in April 2012 by U.S. energy information administration (EIA), the residential electricity usage in U.S. in 2011 is 1,423,700 million kWh consisting 38% of the total electricity energy consumption [1]. This creates an opportunity for residential consumers to play an increasingly active role in DR programs to maintain the balance between electricity demand and supply.

Methods for engaging customers into DR efforts include price-based DR programs via time-varying price mechanisms such as time-of-use pricing, critical peak pricing, variable peak pricing, and real-time pricing, as well as incentive-based intelligent load control DR programs for providing LSEs abilities to cut down/shift loads from peak periods to off-peaks [2]. Meanwhile, with increasing amount of market products and research prototypes [3-13] on home energy management system coming out, realizing intelligent controls largely over residential appliances will become easily feasible in the near future, which would bring tons of benefits to both residents and LSEs. Hence, this paper focuses on incentive-based load control DR programs.

Various examples of algorithms and techniques for optimally scheduling residential demand in DR programs have been discussed. Ref [14] and [15] proposed models to control the aggregated demand from a population of ACs, through the adjustment of temperature set points. And, [16] proposes a method to characterize the availability of residential appliances to provide reserve services with considering the information, such as consumption habits and comfort patterns, 24 hours in advance. However, on the one hand, the previous literatures rarely consider the practical issues with realizing residential demand aggregation such as generating optimal schedules for a large number of appliances in real time. On the other hand, existing literatures seldom discuss about the rewarding system to fairly distribute the rewards (financial incentives) to the participants. And, this issue is essential, because it may affect the residents' participation levels directly.

Therefore, this paper proposes an optimal framework for RLAs to provide effective demand-side ancillary service by strategically controlling residential appliances and impartially rewarding the DR program participants. In the proposed design, each RLA serves as an agent, who receives demand response requests (DRRs) from LSEs and real-time environmental parameters from every household as shown in Fig.1, to optimize the schedule of loads based on the individual residents' preferences, and then send out the optimized control strategies to the actual appliances.

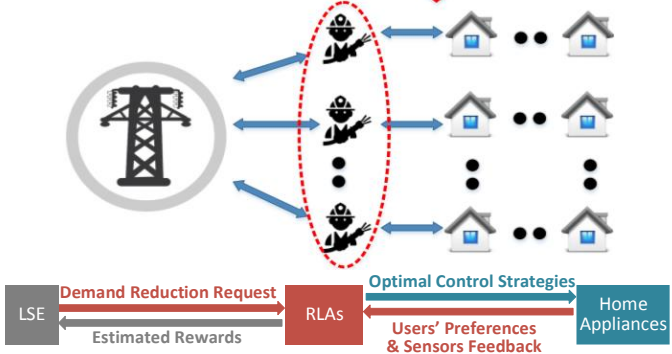

Fig. 1 schematic information flow chart of the optimal framework

In the proposed framework design, to residents, RLAs are able to 1) fairly distribute financial rewards, and 2) maintain residents' comfortableness based on their personal preferences; To LSE, RLAs are able to 1) realize the DRR by controlling residents' appliances, and 2) minimize the total reward costs for performing the DRR. Most importantly, RLAs have fair and strategic rewarding system which enables the residents to become more active and able to customize their energy usage preferences. This may stimulate the potential capability of residential appliances optimized and controlled by RLAs.

The rest of this paper is organized as follows. Section II presents the overall structure of the framework for RLAs. Section III describes the detailed model in math, which is formulated into a mixed-integer quadratically constrained program (MIQCP) problem to minimize the total reward payment while maximizing the residents' comfortableness. The simulation results and numerical analysis of both the small ten residents' system case study and large scale system test are presented in Section IV. The summary and conclusion come in Section V.

## II. OVERVIEW OF THE OPTIMAL FRAMEWORK

Based on several pilot trial runs by utilities[17]-[22], ACs and EWHs are especially critical loads in residential aspect, because they are increasingly predominant and can provide fast responses with minimal impact to residents in a short time period. Moreover, in residential aspects, ACs and EWHs typically account for more than one half of the total peak demand [23]. Therefore, this paper starts from considering the demand aggregation of populations of ACs and EHWs, and plans to integrate electrical vehicles, energy storage components and distributed renewables generation in the future.

There are several assumptions for the proposed framework: 1) ACs and EWHs have bi-directional communication with RLAs, it also indicates that RLAs are able to obtain the real time room temperature and the water tank temperature of EWHs from these appliances; 2) The real-time ambient temperature is known to RLAs; 3) Residents provide comfortable temperature ranges of both room and hot water to RLAs; and 4) Residents decide whether they are willing to compromise, if RLAs have to adjust(lower) the comfortableness of some residents. Following the above assumptions, the RLA will be able to apply proposed framework to dispatch DRRs without affecting residents' normal life, while fairly rewarding residents according to the contributions they made.

Figure 2 is a brief schematic diagram of the control and rewards mechanism. When the RLA receives the request from the LSE saying there is demand $D_r$ need to be reduced, this RLA considers residents' appliances profile, their preferences, ambient temperature, and the real time environmental parameters from the sensors on ACs and EWHs, then will perform the optimization within a very short time. As a result, the RLA achieved several tasks: 1) generated and sent out the optimal control instructions to residents' appliances; 2) provided the LSE with an cost-effective way of realizing the DRR with minimal reward costs; 3) recorded the contributions that individual residents made for this DRR; and 4) distributed the financial rewards to the residents.

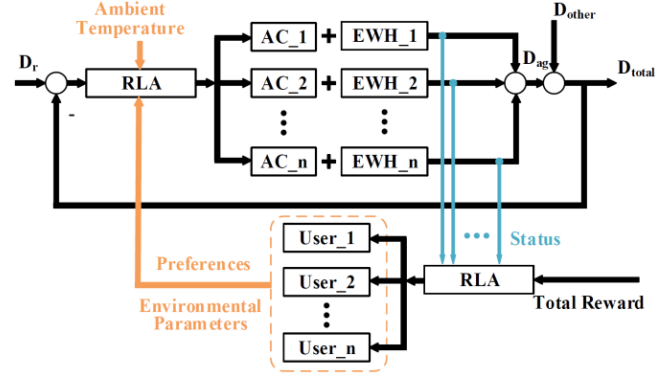

Fig. 2 brief schematic diagram of the control and rewards mechanism

An appealing rewarding system is one of the most important factors to make the optimal framework feasible and then provide effective demand-side ancillary services. The rest content of the section discusses the details on how the rewarding system ensures its effectiveness and fairness.

### A. Rewarding System
#### 1) Various Reward Rates

In this optimal framework, DR program participants will be rewarded fairly and strategically. In order to get rewards, participants have to provide their preferences including the comfortable setting interval for EWH and AC units, and whether they are willing to compromise with turning off EWH and AC unit even if either the room or the water tank temperature will go beyond their comfortable ranges. Whenever the RLA has to make some residents compromise, those residents will receive extra compensation which means higher reward rate for participating in such DRR. Moreover, if there is an emergency issue occurring, in order to maintain the stability of the power system, the LSE has to send a DRR with tremendous amount to RLA. Then, the RLA figured that executing such DRR will have to make the appliances of residents, who claim not to compromise, operate beyond their comfortable ranges. In this case, those participated residents will get the highest reward rate.

Generally, the differences among various reward rates are as shown in Table I. And, take AC units for example, the reward rates for resident $i$ can be determined based on the flow chart as shown in Fig. 3. Mathematically, the various reward rates can be expressed as (1). Since the optimal framework minimizes the total reward costs for LSEs, naturally, the higher the reward level is, the less probability such situation happens. (The minimization of total reward costs will be discussed in section III.)

TABLE I. VARIOUS REWARD RATES

| Resident Type | Rate | Symbol | Probability |
|---|---|---|---|
| Compromise | Common | $R_1$ | Common |
|  | Higher | $R_2$ | Occasional |
| Not Compromise | Common | $R_1$ | Common |
|  | Highest | $R_3$ | Emergency/ Scheduled Maintenance |

$$RWR_{A,i} = \begin{cases} R_1 & ,if \quad T_{L,i} \leq T_{RM,i} \leq T_{H,i} \\ R_2 & ,if \quad T_{RM,i} \leq T_{L,i} \text{ and } Cop_i = 1 \\ R_2 & ,if \quad T_{H,i} \leq T_{RM,i} \text{ and } Cop_i = 1 \\ R_3 & ,if \quad T_{RM,i} \leq T_{L,i} \text{ and } Cop_i = 0 \\ R_3 & ,if \quad T_{H,i} \leq T_{RM,i} \text{ and } Cop_i = 0 \end{cases} \quad (1)$$

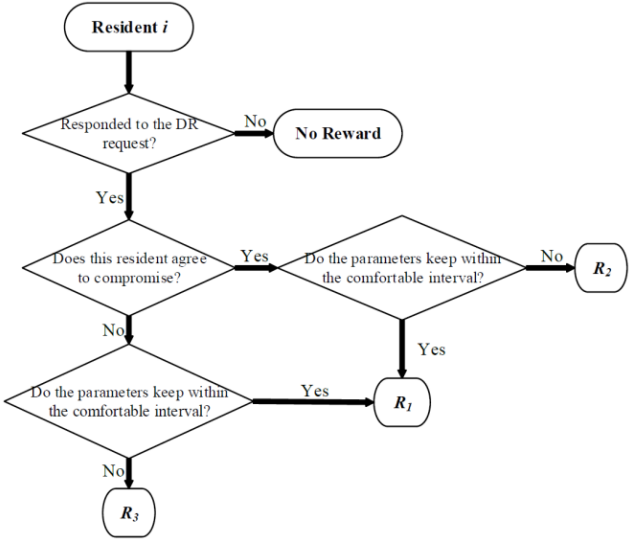

Fig.3 The flow chart of determining the reward rates for resident $i$

*2) Comfortable Margin*

While allocating DRRs to appliances, the RLA is always facing the issue that how to select the proper available appliances to turn off, and how to achieve fairness and justice among all residents. Here, the proposed framework introduces the concept of "comfortable margin" to solve this issue. Take resident $i$ with an AC unit as example, the "comfortable margin", $CM_i$, is defined as equation (2). And the mean value of the low and high threshold of the comfortable range $\frac{T_{L,i}+T_{H,i}}{2}$ is assumed as the perfect operating point. Then, $CM_i$ stands for the distance between the present status and the perfect operating point. Therefore, the higher the $CM_i$ value is, the less comfortable the resident $i$ feel. When temperature goes beyond the comfortable range, $CM_i > 1$.

$$CM_i = \left| \frac{2T_{RM,i} - T_{L,i} - T_{H,i}}{T_{H,i} - T_{L,i}} \right| \quad (2)$$

Substituting (2) into (1), the relationship between $CM_i$ and reward rates can be expressed as (3).

$$RWR_{A,i} = \begin{cases} R_1 & ,if \quad CM_i \leq 1 \\ R_2 & ,if \quad CM_i > 1 \text{ and } Cop_i = 1 \\ R_3 & ,if \quad CM_i > 1 \text{ and } Cop_i = 0 \end{cases} \quad (3)$$

To be fair to all the residents under the control of one RLA, whenever the RLA receives a DRR, it should try to maintain the comfortable margin of each resident similar in the controlled area while performing the DR. This issue has been considered in the objective function of the optimization problem formulation and can be automatically ensured. However, there is still an issue among the residents with same $CM$ values. In the proposed framework, the RLA keeps a record on the history of the DRR participation for every resident. This way, the RLA will be able to choose the one with lower contribution records to participate in maintaining the justice. For example, if the contribution history is as shown in Fig.4 for all the residents, resident#2 and #3 have the same $CM$ value, and one of resident#2 or #3 need to perform a demand reduction on AC unit. According the above rules, resident#2 will be selected.

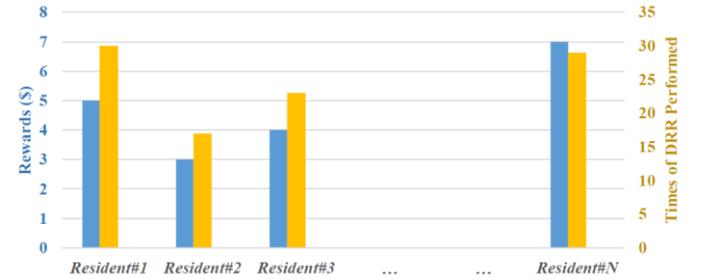

Fig. 4 The record of DRR participation history

Furthermore, it needs to be noted that the record of DRR participation history and rewards distribution results will be kept in the RLA database. Those data will only be uploaded to the LSE every week or month which also help release the stress for LSE to have massive real-time bi-direction communication with tens of thousands of residents.

*B. Analysis of Residents' Strategy*

This optimal framework provides a chance for residents to gain financial rewards in terms of participating DR program. As for a good design, it must be able to attract more program participants while preventing a few malicious manipulation. The proposed framework provides such a platform that satisfies various residents with different needs. By customizing their preferences, residents can involve the DR program at different levels.

Here, a simple example of resident A, B and C with AC units under one RLA is used to perform a general analysis on different residents' strategies without performing detailed optimization calculation. The preferences settings of AC units for A, B, and C are shown in Table II as follow. The comfort intervals of B and C are broader than A's; C chooses to compromise while A and B select not to.

TABLE II. PREFERENCES SETTINGS FOR A, B AND C

| Resident Name | Room Temperature (°F) | Compromise? |
|---|---|---|
| A | 73-77 | No |
| B | 70-80 | No |
| C | 70-80 | Yes |

This example assumes that it is a summer time, resident A, B and C have exactly the same houses and AC units, and the present room temperature is the perfect operating point as mentioned in subsection II-A-2 (A:75 °F, B:75 °F, and C:75 °F). Hence, according to the description, Fig.5 shows the situations of the three residents. The blue curve is rewards rates they will get with different predicted room temperature during the demand reduction period, and the red dotted line is initial room temperature before a DRR.

Now, assume the RLA receives a DRR asking for reducing one third of the total residential demand. Therefore, the RLA needs to shut one of the ACs from these three residents. In CASE#1, because of the same houses, same AC units, and same initial room temperature, the estimated room temperature with executing this DRR is predicted as 77 °F for all three residents, where the reward rate are the same for them. However, due to the concern of *CM*, resident A is excluded, while B and C share the same possibilities to perform the DRR. In CASE#2, the estimate room temperature with executing this DRR is estimated as 81 °F for all three residents, where the reward rate for C is lower than A and B. Therefore, C will be selected to perform the DRR.

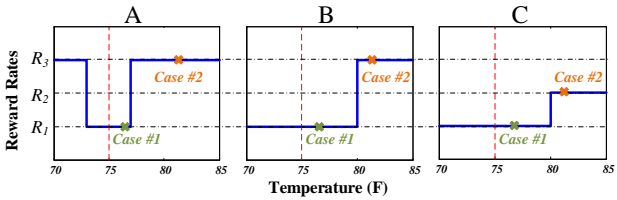

Fig.5 rewards rate with predicted estimate room temperatures for A, B and C

In sum, this simple example shows clearly that resident C will most probably get the chance to perform a demand reduction, and gain financial rewards. Because resident C has the broadest comfortable temperature range setting and is willing to compromise, which means C sacrifices her/his comfortableness most.

This simple example only generally demonstrates how the system works with residents' preferences in brief. However, the practical cases will be much more complex by considering differences in houses, appliances parameters, initial room temperatures, and etc. The next section will provide the complete mathematical formulation of the optimization problem.

## III. OPTIMIZATION PROBLEM FORMULATION

The framework is proposed to realize cost-effective DRR while trying to maintain the comfortableness of residents. In formulating this complete mathematical model, there are several issues with the time length of DRR, temperature estimation, demand reduction accuracy and etc. This section will discuss these issues first and then formulate the complete optimization question into an MIQCP problem which is solvable using available optimization software.

### A. Time Length of DRR

A DRR contains two important information: 1) how much demand to be reduced is; and 2) how long the demand reduction should last. In order to prevent the uncomfortableness caused by performing one single DRR with long time length, the time length of each DRR is set to be five minutes which means the long time length DRR will be treated as several continuous short DRRs.

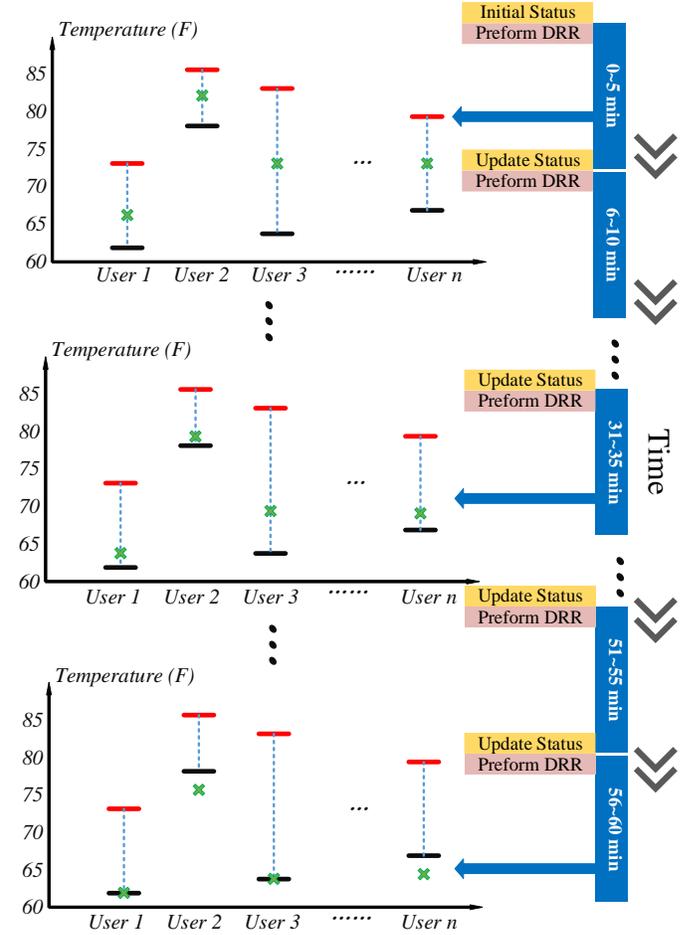

Fig. 6 the process chart for performing one hour length DRR

There are several other advantages with dividing long DRR into short ones. This method ensures a stable calculation time and make the online optimization possible, because it keeps the size of the optimization problem same. Moreover, short DRRs reduce the errors in estimating temperature compared with longer term perdiction, because the sensors' feedbacks will help correct the estimation.

Taking a one-hour long DRR with only ACs at winter time as an example, it will be divided into twelve five-minute

DRRs. After performing each short DRR, the RLA will receive the updated environmental data from the sensors, and then perform the next short DRR after 5 minutes. As the schematic process chart shown in Fig. 6, such method help perform DRR in real-time well and maintain the residents' comfortableness fairly even under long time length DRR.

### B. Temperature Estimation

Temperature estimation is vital in this model, because it determines the reward rates for the residents.

As for EWHs, the general model has been discussed in [24]–[26]. The discrete state dynamics model is applied here, since the time length of each DRR is set to be five minutes. The model can be described by (4):

$$T_{T,i} = -LR_{T,i} \cdot (T0_{T,i} - TA_i) + E_i \cdot PE_i \cdot SE_i \quad (4)$$

As for modelling AC units, the American society of heating, refrigeration and air conditioning engineers, inc. (ASHRAE) has compiled modeling procedures in its fundamentals handbook [27]. The department of energy (DOE) has produced the Energy Plus program for computer simulation [28]. Also, the detailed model for simulating AC systems is given in [29], [30]. According to these literatures, the accurate model needs to consider many factors including weather, season, thermal resistance of rooms, solar heating, cooling effect of the wind, and shading. Unlike EWH which has constant and relatively accurate parameters, those AC parameters are difficult to be precisely modeled since they are always changing with the operating status. In the proposed framework, because of the following reasons: 1) the errors of predicting the temperature for only five minutes ahead is limited, 2) the AC model cannot be complex since this framework needs to perform online optimization. Hence, the simplest model, similar to the model of EWH, as shown in (5) has been implemented.

$$T_{RM,i} = -LR_{RM,i} \cdot (T0_{RM,i} - TA_i) + AE_i \cdot PA_i \cdot SA_i \quad (5)$$

Instead of executing complicated setting adjustments for ACs, in (5), the control variable for AC model is binary $SA_i$. Therefore, the ACs will be controlled by simply ON/OFF.

It has to be highlighted that the values of parameters $LR_{T,i}$, $E_i$, $LR_{RM,i}$, and $AE_i$ are different for every resident. Because the RLA is able to receive the feedbacks of the sensors, the values of those parameters can be obtained through performing regression on the historical data for each resident in practical application in the proposed framework. However, as for the numerical case studies in section IV, due to the lack of historical data, those parameters are only set by assumptions.

### C. Demand Reduction Accuracy Relaxation

Since control strategies for ACs and EWHs in the proposed design are both the arrays with ON/OFF, the total demand can be reduced, expressed as (6).

$$TDR = \sum_{i=1}^{n} PA_i \cdot (1 - SA_i) + PE_i \cdot (1 - SE_i) \quad (6)$$

However, since $SA_i$ and $SE_i$ are both binary, TDR and the demand requested to reduce D, usually cannot be exactly the same since D might go beyond the capability of the RLA. Therefore, the constraint of the amount of demand to be reduced needs to relax according to the LSE's requirement as (7). And, the value of $\delta$ is set as 0.05 in the following case study in subsection IV-2.

$$(1-\delta) \cdot D \leq TDR \leq (1+\delta) \cdot D \quad (7)$$

### D. MIQCP Problem Formulation

Given the discussions above, this optimization problem of minimizing total rewards payment while maximizing the residents' comfortableness during the summer time can be formulated as:

$$\min \sum_{i=1}^{n} RW_i + w \cdot \sum_{i=1}^{n} CM_i^2 \quad (8a)$$

s.t.

$$RW_i = PE_i \cdot (1 - SE_i) \cdot RWR_{E,i} + PA_i \cdot (1 - SA_i) \cdot RWR_{A,i} \quad (8b)$$

$$TDR = \sum_{i=1}^{n} PA_i \cdot (1 - SA_i) + PE_i \cdot (1 - SE_i) \quad (8c)$$

$$(1-\delta) \cdot D \leq TDR \leq (1+\delta) \cdot D \quad (8d)$$

$$T_{T,i} = -AE_i \cdot (T0_{T,i} - TA_i) + E_i \cdot PE_i \cdot SE_i \quad (8e)$$

$$T_{RM,i} = -LR_i \cdot (T0_{RM,i} - TA_i) + AE_i \cdot PA_i \cdot SA_i \quad (8f)$$

$$RWR_{A,i} = \begin{cases} R_1, & \text{if } T_{RM,i} \leq T_{H,i} \\ R_2, & \text{if } T_{H,i} \leq T_{RM,i} \text{ and } Cop_i = 1 \\ R_3, & \text{if } T_{H,i} \leq T_{RM,i} \text{ and } Cop_i = 0 \end{cases} \quad (8g)$$

$$RWR_{E,i} = \begin{cases} R_1, & \text{if } T_{TL,i} \leq T_{T,i} \\ R_2, & \text{if } T_{TL,i} \geq T_{T,i} \text{ and } Cop_i = 1 \\ R_3, & \text{if } T_{TL,i} \geq T_{T,i} \text{ and } Cop_i = 0 \end{cases} \quad (8h)$$

$$CM_i = \left| \frac{2T_{RM,i} - T_{L,i} - T_{H,i}}{T_{H,i} - T_{L,i}} \right| + \left| \frac{2T_{T,i} - T_{TL,i} - T_{TH,i}}{T_{TH,i} - T_{TL,i}} \right| \quad (8i)$$

In order to make this problem solvable, the constraint (8g) can be converted as (9a), (9b) and (9c).

$$RWR_{A,i} = R_1 \cdot \mu_i + R_2 \cdot (1-\mu_i) \cdot Cop_i + R_3 \cdot (1-\mu_i) \cdot (1-Cop_i) \quad (9a)$$

$$T_{RM,i} - T_{H,i} \leq M \cdot (1-\mu_i) \quad (9b)$$

$$T_{RM,i} - T_{H,i} > -M \cdot \mu_i \quad (9c)$$

Similarly, the constraint (8h) can be converted to (10a), (10b) and (10c).

$$RWR_{E,i} = R_1 \cdot \upsilon_i + R_2 \cdot (1-\upsilon_i) \cdot Cop_i + R_3 \cdot (1-\upsilon_i) \cdot (1-Cop_i) \quad (10a)$$

$$T_{TL,i} - T_{T,i} \leq M \cdot (1-\upsilon_i) \quad (10b)$$

$$T_{TL,i} - T_{T,i} > -M \cdot \upsilon_i \quad (10c)$$

where M is large enough constants, and $\mu_i$ and $\upsilon_i$ are the auxiliary binary variables [31].

Therefore, the optimization problem is formulated as a MIQCP problem, which is easy to solve by the available software, as below.

*min* (8a)

*s.t.*

constraints (8b), (8c), (8d), (8e), (8f), (8i), (9a), (9b), (9c), (10a), (10b), (10c).

## IV. CASE STUDIES

In this section, the proposed optimal framework is performed on both a ten residents' system and a five hundred residents system which is based on the residential energy consumption survey (RECS) made by U.S. EIA in 2009 [32]. The first case study is supposed to show the rewarding system for each resident, since the small case study will demonstrate more detailed information. Further, the second case study is used to show the changes of residents' comfortableness and total rewards costs for the LSE under different DRRs under the proposed framework.

The simulation has been done in General Algebraic Modeling System (GAMS) which has the capability to solve large scale optimization problems. The MINQCP problem is solved by BONMIN solver in GAMS on a desktop with Intel Xeon 3.2GHz CPU, 8 GB RAM, and Window 8.

### A. Ten Residents' System

Based on the proposed framework and optimization problem formulation, several case studies have been carried out. The first test system is a ten residents' system considering only AC units. In this system, every residents has different personal preferences and house household parameters as shown in Table III. The total regular demand of ACs is 13.6 kW, and reward rates $R_1$, $R_2$, and $R_3$ are 10, 20, and 30 cents/ (kW·5min) respectively.

TABLE III. TEN RESIDENTS' PROFILE

| ID | $T_H$ | $T_L$ | PA | T0 | Cop | AE | LR |
|---|---|---|---|---|---|---|---|
| 1 | 75 | 70 | 1.3 | 72.5 | 0 | 5 | 0.1 |
| 2 | 75 | 70 | 1.4 | 72.5 | 1 | 5 | 0.1 |
| 3 | 75 | 65 | 1.2 | 70 | 0 | 5 | 0.3 |
| 4 | 80 | 70 | 1.5 | 75 | 0 | 5 | 0.2 |
| 5 | 75 | 65 | 1.6 | 70 | 1 | 6 | 0.3 |
| 6 | 75 | 65 | 1.3 | 70 | 1 | 5 | 0.1 |
| 7 | 75 | 67 | 1.2 | 71 | 0 | 4 | 0.1 |
| 8 | 77 | 67 | 1.1 | 70 | 1 | 4 | 0.2 |
| 9 | 77 | 65 | 1.5 | 71 | 0 | 5 | 0.2 |
| 10 | 75 | 70 | 1.5 | 72.5 | 1 | 5 | 0.2 |

In this ten residents' case study, the RLA received two DRRs from the LSE.

*1) DRR#1 with 4kW/ 20min*

The DRR#1 asks the RLA to reduce 4kW for 20 min among these ten residents' AC units. The results of residents' satisfaction as well as rewards distribution are as shown in Table IV. CMFT stands for the percentage of the time when the temperature is within the comfortable ranges.

As for the results of DRR#1, all the residents are within their comfortable temperature ranges. Resident #6 gets the most financial rewards, due to his board comfortable temperature range and low LR value. The lower LR means the lower temperature change while turning off the appliances. In other words, low LR enhances the capability for the resident in participating DRR. Neither resident #3 nor #5 has rewards, because they have to keep their ACs on to maintain the proper room temperature due to the high LRs. As a result, RLA will not try to turn their ACs off, while others can offer enough demand reduction.

TABLE IV. DRR#1 RESULT

| ID | min $T_{RM}$ (°F) | max $T_{RM}$ (°F) | CMFT (%) | Rate | Rewards ($) |
|---|---|---|---|---|---|
| 1 | 70.8 | 74.5 | 100 | $R_1$ | 1.9 |
| 2 | 70 | 74.3 | 100 | $R_1$ | 2.8 |
| 3 | 71 | 72.8 | 100 | $R_1$ | 0 |
| 4 | 70.1 | 74.1 | 100 | $R_1$ | 1.5 |
| 5 | 70 | 70 | 100 | $R_1$ | 0 |
| 6 | 66.4 | 70.1 | 100 | $R_1$ | 3.9 |
| 7 | 70.3 | 72.9 | 100 | $R_1$ | 1.2 |
| 8 | 70 | 74 | 100 | $R_1$ | 1.1 |
| 9 | 67 | 69.8 | 100 | $R_1$ | 2 |
| 10 | 70 | 74.8 | 100 | $R_1$ | 1.5 |

*2) DRR#2 with 8kW/ 20min*

The DRR#2 requests the RLA to reduce 8kW for 20 min among these ten residents' AC units. The results of residents' satisfaction as well as reward distribution are as shown in Table V.

TABLE V. DRR#2 RESULT

| ID | min $T_{RM}$ (°F) | max $T_{RM}$ (°F) | CMFT (%) | Rate | Rewards ($) |
|---|---|---|---|---|---|
| 1 | 70.8 | 74.5 | 100 | $R_1$ | 2.9 |
| 2 | 70 | 75.8 | 75 | $R_1, R_2$ | 6.4 |
| 3 | 71 | 72.5 | 100 | $R_1$ | 0 |
| 4 | 73 | 79.1 | 100 | $R_1$ | 3 |
| 5 | 70 | 80.2 | 50 | $R_1, R_2$ | 6.4 |
| 6 | 68.8 | 72.8 | 100 | $R_1$ | 3.9 |
| 7 | 72.2 | 74.6 | 100 | $R_1$ | 3.6 |
| 8 | 72.6 | 76.1 | 100 | $R_1$ | 2.2 |
| 9 | 72.8 | 76.3 | 100 | $R_1$ | 3 |
| 10 | 71 | 76.3 | 75 | $R_1, R_2$ | 6 |

In DRR#2, the demand to be reduced is around 60% of the total regular demand which is tremendous. Therefore, it causes problem that some of the residents have to bear the hot weather such as resident#2, #5 and #10. Consequently, they get financial reward relatively higher than others, because they are rewarded with $R_2$ when their room temperature goes beyond their comfortable ranges. It needs to note that all these three residents selected willing to compromise.

Comparing the residents' profile and the results of these two DRRs, it seems that all the participants are fairly considered regarding their preferences and parameters. And, the calculation time for performing both DRRs is within 0.02s. Moreover, Table. VI shows clearly that the increase of DRR leads to the dramatic rise, in terms of the reward costs, since the amount of DRR2 is twice as DRR1, but the total reward cost to perform DRR2 is about 2.34 times DRR1 as shown in Table VI.

TABLE VI. RESULT COMPARISON BETWEEN DRR#1 AND #2

| | Duration (min) | Amount (kW) | Ratio to Regular Demand (%) | Average CMFT (%) | Rewards ($) |
|---|---|---|---|---|---|
| DRR1 | 20 | 4 | 29.4% | 100% | 15.9 |
| DRR2 | 20 | 8 | 58.8% | 90% | 37.4 |

Peak electricity demands occurs on hot summer afternoons when air conditioners are working hard to keep homes and business cool. This increased demand is a challenge for power companies and can result in higher costs for power companies and higher bills for the customer.

## B. Five Hundred Residents' System

The data of this five hundred residents' system is from the RECS by U.S. EIA. The RECS data sets collect the information related to the appliances the residents owe, and the parameters as well as usual settings of those appliances. The original RECS contains the information from more than 60,000 households. In this case study, only five hundred are selected, because the RLA is supposed to solve practical problem with this scale.

The result turns out performing the optimization of five hundred residents system take less than 80 seconds in calculation time. And, the change of residents' comfortableness and total rewards costs for the LSE under different DRR are as shown in Fig.7 and Fig.8. The results are reasonable that, with the increase of the time length and the amount of the demand to be reduced in DRR, the residents' comfortableness are dramatically falling while the total reward costs are rising sharply.

(Not only for validating the proposed framework, the result of this case study might provide valuable information for the LSEs who are using incentive-based intelligent load control DR programs.)

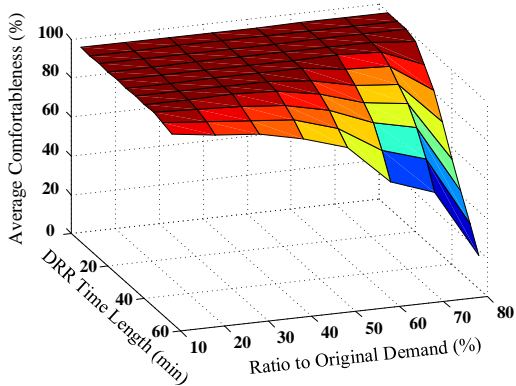

Fig.7 Residents' comfortableness with different DRR

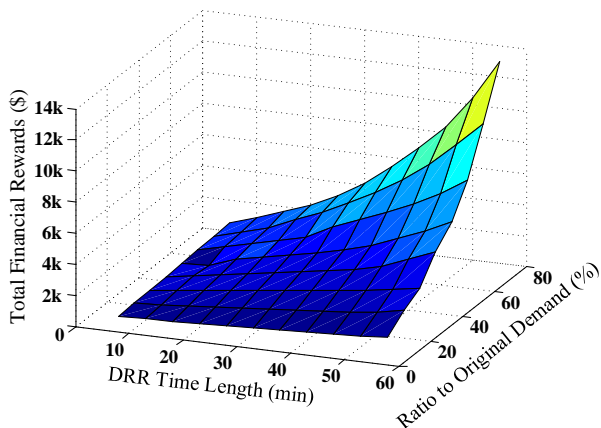

Fig. 8 Total rewards costs for the LSE with different DRR

## V. CONCLUSION

In this paper, an optimal framework for RLAs is proposed. Under this framework, the load aggregators will serve as agents of LSEs to not only allocate DDRs among residential appliances quickly and efficiently without affecting residents' comfortableness, but also fairly and strategically reward residents for their participation, which may stimulate the potential capability of loads optimized and controlled by RLAs in DR program. The main contributions of this framework can be summarized as below:

1) For the LSE, RLAs reduce the size of the optimization problem and make dispatching DRR down to residential appliances feasible in real time.

2) This framework minimizes the total reward costs for LSEs to perform an efficient DRR in demand response program while maintaining the comfortableness for residents.

3) The rewarding system is established to satisfy the needs for various types of electricity consumers. They can make a tradeoff between financial rewards and living comfortableness by strategically and simply setting their preferences over the appliances.

4) Moreover, since this framework both benefits LSEs and appeals to residents, it may stimulate the potential capability of residential appliances optimized and controlled by RLAs in DR program. Eventually, with the growing electricity usage in residential aspect, this framework will have the opportunity to become one of the most vital part in providing effective demand-side ancillary services for the whole power system.

The future work of present framework includes integrating the models of electrical vehicle and energy storage component, and considering the possibilities of bi-directional electricity transfer between LSEs and common residents with distributed generation devices installed.